\documentclass
[superscriptaddress,secnumarabic,amssymb,amsmath,nobibnotes,aps,prd,nofootinbib,nopacsnumber,onecolumn,12pt]{revtex4}%
\usepackage{placeins}
\usepackage{graphicx}
\usepackage{epsf}
\usepackage{bm}
\usepackage{amsmath}
\usepackage{amsfonts}
\usepackage{amssymb}%
\providecommand{\U}[1]{\protect\rule{.1in}{.1in}}

\newcommand{\be}{\begin{equation}}
\newcommand{\ee}{\end{equation}}

\newcommand{\mincir}{\raise
-3.truept\hbox{\rlap{\hbox{$\sim$}}\raise4.truept\hbox{$<$}\ }}
\newcommand{\magcir}{\raise
-3.truept\hbox{\rlap{\hbox{$\sim$}}\raise4.truept\hbox{$>$}\ }}

\RequirePackage{latexsym}
\usepackage{hyperref}
\hypersetup{
    colorlinks=true,
    linkcolor=red,
    citecolor=blue,
    filecolor=magenta,      
    urlcolor=cyan,
    pdftitle={Averaging generalized scalar field cosmologies IV: locally rotationally symmetric Bianchi V model},
    pdfpagemode=FullScreen}  
\newcommand{\bea}{\begin{eqnarray}}
\newcommand{\eea}{\end{eqnarray}}

\newcommand{\bef}{\begin{figure}}  \newcommand{\eef}{\end{figure}}
\newcommand{\bec}{\begin{center}}  \newcommand{\eec}{\end{center}}

\usepackage{url} 
\usepackage{pdflscape} 
\usepackage{amsmath}
\usepackage{amssymb} 
\usepackage{epsfig} 
\usepackage{amsfonts}
\usepackage{graphicx}
\usepackage{subfigure}
\usepackage{graphicx}
\usepackage{times,fancyhdr}
\usepackage{enumitem}
\setlist[enumerate,2]{label=\roman*)}

\def\case#1/#2{\textstyle\frac{#1}{#2}}

\newcommand{\ben}{\begin{eqnarray}}
\newcommand{\een}{\end{eqnarray}}
\usepackage{mdframed}
\usepackage{grffile}
\usepackage{amssymb}
\usepackage[T1]{fontenc}

\usepackage{amssymb,amsmath,amsfonts,amsthm,graphicx, graphics, setspace}
\usepackage{bigints}
\usepackage{anysize}
\usepackage{float}
\usepackage{fancyhdr}
\usepackage{subfigure}
\usepackage{pdflscape}
\usepackage{epsfig}
\usepackage{epsf}
\usepackage{epstopdf}
\usepackage{hyperref}

\providecommand{\U}[1]{\protect\rule{.1in}{.1in}}

\usepackage{tikz,xcolor,hyperref}

\hypersetup{
    colorlinks=true,
    linkcolor=red,
    citecolor=blue,
    filecolor=magenta,      
    urlcolor=cyan,
    pdftitle={Bianchi I spacetimes in f(Q)-gravity},
    pdfpagemode=FullScreen}  

\definecolor{lime}{HTML}{A6CE39}
\DeclareRobustCommand{\orcidicon}{%
	\begin{tikzpicture}
	\draw[lime, fill=lime] (0,0) 
	circle [radius=0.16] 
	node[white] {{\fontfamily{qag}\selectfont \tiny ID}};
	\draw[white, fill=white] (-0.0625,0.095) 
	circle [radius=0.007];
	\end{tikzpicture}
	\hspace{-2mm}
}

\foreach \x in {A, ..., Z}{%
	\expandafter\xdef\csname orcid\x\endcsname{\noexpand\href{https://orcid.org/\csname orcidauthor\x\endcsname}{\noexpand\orcidicon}}
}

\begin{document}
\title{Bianchi I spacetimes in $f\left(Q\right)$-gravity}
\author{Andronikos Paliathanasis\orcidD{}}
\email{anpaliat@phys.uoa.gr}
\affiliation{Institute of Systems Science, Durban University of Technology, Durban 4000,
South Africa}
\affiliation{Departamento de Matem\'{a}ticas, Universidad Cat\'{o}lica del Norte, Avda.
Angamos 0610, Casilla 1280 Antofagasta, Chile}
\author{Genly Leon\orcidA{}}
\email{genly.leon@ucn.cl}
\affiliation{Departamento de Matem\'{a}ticas, Universidad Cat\'{o}lica del Norte, Avda.
Angamos 0610, Casilla 1280 Antofagasta, Chile}
\affiliation{Institute of Systems Science, Durban University of Technology, Durban 4000,
South Africa}

\begin{abstract}
The Kasner spacetimes are exact solutions that are self-similar and of significant interest because they can describe the dynamic evolution of kinematic variables near the singularity of the Mixmaster universe. The chaotic behavior of the Mixmaster universe is related to the Kasner relations. This study investigates Kasner and Kasner-like solutions in symmetric teleparallel $f\left( Q\right) $-gravity. We consider three families of connections for the spatially flat Friedmann-Lema\^{\i}tre-Robertson-Walker. We found that Kasner-like solutions are present in the theory when the nonmetricity scalar $Q$ vanishes. Kasner-like relations are introduced only for the field equations of the connection defined in the coincidence gauge. Finally, we perform a detailed phase-space analysis to understand the dynamics of $f\left( Q\right) $-gravity in the evolution of anisotropies and the effects of different connections on this evolution.

\end{abstract}
\maketitle

\section{Introduction}

\label{sec1}

Modified theories offer a different way to address the dark energy problem, suggesting that the universe's acceleration is caused by changes to Einstein's General Relativity (GR). GR is a gravitational theory that uses a metric tensor and the Levi-Civita connection to describe physical space. In this theory, the Ricci scalar $\overset{o}{R}$ for the Levi-Civita connection acts as the Lagrangian.

Teleparallel Equivalence of GR (TEGR) \cite{ein28} and Symmetric Teleparallel
GR (STGR) \cite{ngr2} are two different gravitational theories which are
equivalent to GR. In TEGR the geometry is described Weitzenb{\"{o}}ck
connection \cite{Weitzenb23}, in which the corresponding dynamical fields are
the four linearly independent vierbeins and the gravitational Lagrangian is
related to the torsion scalar $T~$related to the antisymmetric connection of
the non-holonomic basis \cite{Hayashi79,Tsamp}. On the other hand, in STGR, the
physical space is described by the metric tensor and a symmetric and flat
connection, which inherits the symmetries of the background space \cite{ngr2}.
The gravitational Action Integral in STGR depends on the nonmetricity scalar
for the connection.

The fundamental scalars of the three equivalent theories of gravity differ by
a boundary term \cite{Hayashi79}, from where it is easy to construct one
gravitational theory from the other using proper boundary corrections.
Nevertheless, the equivalency is lost when nonlinear terms of the scalars are
introduced in the gravitational Lagrangian. The family of $f-$theories of
gravity has been widely studied in the literature as modified theories of GR
\cite{Buda} and TEGR \cite{Ferraro, Ferraro06}. It has been found that these
modified theories of gravity can describe various epochs of the cosmological
history, for instance,
\cite{rr1,rr2,rr3,rr4,rr5,rr6,rr7,rr8,rr9,rr10,rr11,rr12} and references therein.

Recently, there has been a direction in the literature in nonmetricity
$f\left(Q\right)$-theory \cite{fqq1,fqq2,fqq3,fqq4}. In terms of
cosmology, it has been found that $f\left(Q\right)$-theory can explain the
cosmological observations \cite{ds00,ww0,ww8}, and it can describe the main
eras of the cosmological history \cite{dm7,dm8}. Isotropic and homogeneous
cosmological exact solutions in the framework of $f\left(Q\right)$-theory
investigated before in \cite{cos1,cos2,cos3}. A main characteristic of the
$f\left(Q\right)$-theory is that the symmetric and flat connection is not
necessarily unique. For instance, for the spatially flat
Friedmann--Lema\^{\i}tre--Robertson--Walker (FLRW), there are three families of
connections \cite{Hohmann, Heis2}. Recent studies on the cosmological
observations indicate the pathology of $f\left(Q\right)$-cosmology in
describing the matter epoch \cite{per1,per2}. These problems may overlap in
the presence of a chameleon mechanism. However, $f\left(Q\right)$-theory
is a great toy model for the description of the late- or early-time
acceleration phase of the universe \cite{ff1,ff2}.

In this study, we are interested in the derivation of exact Bianchi I
spacetimes in $f\left(Q\right)$-theory of gravity. The Kasner universe
\cite{kasner1} is the most common anisotropic exact solution because it can
describe the chaotic behavior as we reach the singularity at the Mixmaster
universe \cite{DemaretKL}. Because of the importance
\cite{zs, bt,bmass,bher, silk, skew, HT,jbBBN} of the Kasner universe and its
generalization, it has been widely studied in modified theories of gravity
\cite{ks1,ks2,ks3,ks4,ks5}. Moreover, we construct another family of solutions
for the Bianchi I geometry. \ In \cite{bb1} the dynamics in the anisotropic Bianchi I spacetime in the coincidence gauge in
$f\left(Q\right)$-gravity was investigated, some exact solutions for the
LRS Bianchi I geometry in the coincidence gauge studied in \cite{bb2}.
However, to our knowledge, there are no studies on the noncoincidence
gauge for the Bianchi I spacetime. For the Kantowski-Sachs geometry and the
LRS Bianchi III geometry exact solutions in $f\left(Q\right)$-theory with
the connection defined in the noncoincident gauge in
\cite{nd1}. 

The structure of the paper is as follows. Section \ref{sec2} presents the basic definitions for the nonmetricity tensor
and of the symmetric teleparallel $f\left(Q\right)$-gravity. We introduce
the Bianchi I line element in Section \ref{sec3}, where we assume the three
families of connections introduced to describe the spatially flat FLRW
geometry for the symmetric and teleparallel connection. For the three
different connections, we derive the gravitational field equations where we
observe that in the limit of isotropization, we recover the field equations of
the FLRW universe. Section \ref{sec4} includes the main results of this
analysis, where we determine exact anisotropic solutions. Specifically, we
investigate Kasner\ (-like) solutions, where we find that these solutions
exist when the nonmetricity scalar vanishes. Moreover, in Section \ref{sec5},
we present the phase-space analysis for the three different sets of field
equations. From this analysis, we study the effects of the different
connections on the evolution of the cosmological anisotropies. Finally, we draw our conclusions in
Section \ref{sec6}.

\section{$f\left(Q\right)$-gravity}

\label{sec2}

In nonmetricity $f\left(Q\right)$-gravity we assume that the
four-dimensional manifold, which describes the physical space, is
characterized by the metric tensor $g_{\mu\nu}$ and symmetric and flat
connection $\Gamma_{\mu\nu}^{\kappa}$ with the property
\begin{equation}
\nabla_{\kappa}g_{\mu\nu}=Q_{\kappa\mu\nu}\equiv\frac{\partial g_{\mu\nu}%
}{\partial x^{\lambda}}-\Gamma_{\;\lambda\mu}^{\sigma}g_{\sigma\nu}%
-\Gamma_{\;\lambda\nu}^{\sigma}g_{\mu\sigma},
\end{equation}
in which $Q_{\kappa\mu\nu}$ is the contorsion tensor, also known as the
nonmetricity tensor \cite{Eisenhart}. \newline

Because the connection $\Gamma_{\mu\nu}^{\kappa}$ is symmetric, there is not
any torsion component, that is,%
\begin{equation}
\mathrm{T}_{\mu\nu}^{\lambda}\equiv\Gamma_{\;\left[  \mu\nu\right]  }%
^{\lambda}=0,
\end{equation}
and the curvature tensor is always zero
\begin{equation}
R_{\;\lambda\mu\nu}^{\kappa}\equiv\frac{\partial\Gamma_{\;\lambda\nu}^{\kappa
}}{\partial x^{\mu}}-\frac{\partial\Gamma_{\;\lambda\mu}^{\kappa}}{\partial
x^{\nu}}+\Gamma_{\;\lambda\nu}^{\sigma}\Gamma_{\;\mu\sigma}^{\kappa}%
-\Gamma_{\;\lambda\mu}^{\sigma}\Gamma_{\;\mu\sigma}^{\kappa}=0.
\end{equation}

The gravitational Action of nonmetricity $f\left(Q\right)$-gravity is
\cite{Hohmann,Heis2}
\begin{equation}
S_{f\left(  Q\right)  }=\int d^{4}x\sqrt{-g}f(Q),
\end{equation}
where $Q$ is the nonmetricity scalar defined as \cite{ngr2}%
\[
Q=Q_{\lambda\mu\nu}P^{\lambda\mu\nu}%
\]
where
\begin{equation}
Q_{\mu}=Q_{\mu\nu}^{\phantom{\mu\nu}\nu}, \bar{Q}_{\mu
}=Q_{\phantom{\nu}\mu\nu}^{\nu\phantom{\mu}\phantom{\mu}}, 
P_{\;\mu\nu}^{\lambda}=-\frac{1}{4}Q_{\;\mu\nu}^{\lambda}+\frac{1}{2}%
Q_{(\mu\phantom{\lambda}\nu)}^{\phantom{(\mu}\lambda\phantom{\nu)}}+\frac
{1}{4}\left(  Q^{\lambda}-\bar{Q}^{\lambda}\right)  g_{\mu\nu}-\frac{1}%
{4}\delta_{\;(\mu}^{\lambda}Q_{\nu)}.
\end{equation}

Let $\overset{o}{R}$ describes the Ricci scalar defined by the Levi-Civita
connection for the metric tensor $g_{\mu\nu}$, it holds that
\begin{equation}
\int d^{4}x\sqrt{-g}Q\simeq\int d^{4}x\sqrt{-g}\overset{o}{R}+\text{boundary
terms,}%
\end{equation}
which means that $f\left(Q\right)$-gravity in the linear case is
equivalent to General Relativity. This theory is known as Symmetric
Teleparallel General Relativity (STGR).

Varying of the Action Integral $S_{f\left(  Q\right)  }$ with respect to the
metric tensor gives the modified field equations \cite{Hohmann, Heis2}
\begin{equation}
\frac{2}{\sqrt{-g}}\nabla_{\lambda}\left(  \sqrt{-g}f^{\prime}(Q)P_{\;\mu\nu
}^{\lambda}\right)  -\frac{1}{2}f(Q)g_{\mu\nu}+f^{\prime}(Q)\left(  P_{\mu
\rho\sigma}Q_{\nu}^{\;\rho\sigma}-2Q_{\rho\sigma\mu}P_{\phantom{\rho\sigma}\nu
}^{\rho\sigma}\right)  =0,
\end{equation}
where $f^{\prime}\left(  Q\right)  =\frac{df}{dQ}$.

On the other hand, variation with respect to the connection gives the equation
of motion
\begin{equation}
\nabla_{\mu}\nabla_{\nu}\left(  \sqrt{-g}f^{\prime}%
(Q)P_{\phantom{\mu\nu}\sigma}^{\mu\nu}\right)  =0.
\end{equation}

The equation of motion is possibly identically satisfied; in this case, we
shall say that the connection is defined in the \textquotedblleft coincidence
gauge\textquotedblright. Recall that the connection has a different
transformation rule from the tensors, meaning it is coordinate-dependent.
Because it is flat, a coordinate system exists where all the components can be
zero. However, we have already defined the coordinate system by selecting a
specific metric tensor. Hence, when the equation of motion for the connection
is not trivially zero for this specific coordinate system, we shall say that
the connection is defined in the \textquotedblleft noncoincidence
gauge\textquotedblright.

\section{Bianchi I geometry}

\label{sec3}

Bianchi I spacetime is the simplest model provided by the Bianchi
classification scheme where the line element is
\begin{equation}
ds^{2}=-dt^{2}+a^{2}(t)dx^{2}+b^{2}(t) dy^{2}+c^{2}(t) dz^{2}, \label{b1}%
\end{equation}
in which $a(t) $,~$b(t) $ and $c(t) $ are the three scale factors. The later
line element admits three isometries, which are the three translations of the
flat space, that is, the vector fields $\partial_{x}$, $\partial_{y}$ and
$\partial_{z}$. When the three scale factors are equal, the Bianchi I geometry admits the
additional isometries, and the spacetime is reduced to that of the spatially
flat FLRW space. Therefore, the three symmetric and flat connections 
describe the FLRW space in nonmetricity theory and the Bianchi I geometry.

In Cartesian coordinates, the nonzero components of the three connections are
\[
\Gamma_{1}:~\Gamma_{tt}^{t}=\gamma(t),
\]%
\[
\Gamma_{2}:~\Gamma_{tt}^{t}=\gamma(t)+\frac{\dot{\gamma}(t)}{\gamma
(t)},~\Gamma_{tx}^{x}=\Gamma_{ty}^{x}=\Gamma_{tz}^{x}=\gamma(t),~
\]%
\[
\Gamma_{3}:~\Gamma_{tt}^{t}=-\frac{\dot{\gamma}(t)}{\gamma(t)},~\Gamma
_{tx}^{t}=\Gamma_{ty}^{t}=\Gamma_{tz}^{t}=-\gamma(t).
\]
Function $\gamma$, as we shall see, introduces dynamic degrees of freedom in
the field equations.

Connection $\Gamma_{1}$ is defined the coincidence gauge while connections
$\Gamma_{2}$ and $\Gamma_{3}$ are defined in the noncoincidence gauge. For
each connection, the gravitational field equations are the same in the case of
STGR, but differs for a nonlinear $f\left(Q\right)$-theory.

\subsection{Field equations for connection $\Gamma_{1}$}

For the connection $\Gamma_{1}$ we calculate the nonmetricity scalar
\begin{equation}
Q\left(  \Gamma_{1}\right)  =-2\left(  H_{a}H_{b}+H_{a}H_{c}+H_{b}%
H_{c}\right)  \label{con.00}%
\end{equation}
where $H_{a}=\frac{\dot{a}}{a}~,~H_{b}=\frac{\dot{b}}{b}$ and $H_{c}%
=\frac{\dot{c}}{c}$.

Therefore, the gravitational field equations are
\begin{align}
f^{\prime}\left(  H_{a}H_{b}+H_{a}H_{c}+H_{b}H_{c}\right)  +\frac{1}{2}\left(
f-Qf^{\prime}\right)  =0, \label{con.01}%
\\
f^{\prime}\left(  \left(  H_{b}+H_{c}\right)  \left(  H_{a}+H_{b}%
+H_{c}\right)  +\dot{H}_{b}+\dot{H}_{c}\right)  +\dot{Q}f^{\prime\prime
}\left(  H_{b}+H_{c}\right)  +\frac{f}{2}=0, \label{con.02}%
\\
f^{\prime}\left(  \left(  H_{a}+H_{c}\right)  \left(  H_{a}+H_{b}%
+H_{c}\right)  +\dot{H}_{a}+\dot{H}_{c}\right)  +\dot{Q}f^{\prime\prime
}\left(  H_{a}+H_{c}\right)  +\frac{f}{2}=0, \label{con.03}%
\\
f^{\prime}\left(  \left(  H_{a}+H_{b}\right)  \left(  H_{a}+H_{b}%
+H_{c}\right)  +\dot{H}_{a}+\dot{H}_{b}\right)  +\dot{Q}f^{\prime\prime
}\left(  H_{a}+H_{b}\right)  +\frac{f}{2}=0. \label{con.04}%
\end{align}

Because $\Gamma_{1}$ is defined in the coincidence gauge, the equation of
motion for the connection is trivially satisfied.

\subsubsection{Minisuperspace description}

The latter gravitational field equations admit a point-like Lagrangian.
Indeed, by introducing the scalar field $\phi=f^{\prime}\left(  Q\right)  $
and the potential function $V\left(  \phi\right)  =\frac{1}{2}\left(
f-Qf^{\prime}\right)  $, it follows that the field equations can be derived
from the variation of the Action Integral%
\begin{equation}
S=\int L_{\Gamma_{1}}\left(  a,\dot{a},b,\dot{b},c,\dot{c},\phi,\dot{\phi
}\right)  dt,
\end{equation}
in which the Lagrangian function $L_{\Gamma_{1}}$ is defined as%
\begin{equation}
L_{\Gamma_{1}}\left(  a,\dot{a},b,\dot{b},c,\dot{c},\phi,\dot{\phi}\right)
=\phi\left( c\dot{a}\dot{b}+b\dot{a}\dot{c}+a\dot{b}\dot{c}\right)
-abcV\left(  \phi\right) .
\end{equation}

\subsection{Field equations for connection $\Gamma_{2}$}

For the second connection, we find
\begin{equation}
Q\left(  \Gamma_{2}\right)  =-2\left(  H_{a}H_{b}+H_{a}H_{c}+H_{b}%
H_{c}\right)  +3\left(  \left(  H_{a}+H_{b}+H_{c}\right)  \gamma+\dot{\gamma
}\right),  \label{con2.01}%
\end{equation}
and the field equations are
\begin{align}
&  f^{\prime}\left(  H_{a}H_{b}+H_{a}H_{c}+H_{b}H_{c}\right)  +\frac{1}%
{2}\left(  f-Qf^{\prime}\right)  +3\gamma\dot{Q}f^{\prime\prime}=0,
\label{con2.02}\\
&  f^{\prime}\left(  \left(  H_{b}+H_{c}\right)  \left(  H_{a}+H_{b}%
+H_{c}\right)  +\dot{H}_{b}+\dot{H}_{c}\right)  +\dot{Q}f^{\prime\prime
}\left(  H_{b}+H_{c}-\frac{3}{2}\gamma\right)  +\frac{f}{2}=0, \label{con2.03}%
\\
&  f^{\prime}\left(  \left(  H_{a}+H_{c}\right)  \left(  H_{a}+H_{b}%
+H_{c}\right)  +\dot{H}_{a}+\dot{H}_{c}\right)  +\dot{Q}f^{\prime\prime
}\left(  H_{a}+H_{c}-\frac{3}{2}\gamma\right)  +\frac{f}{2}=0, \label{con2.04}%
\\
&  f^{\prime}\left(  \left(  H_{a}+H_{b}\right)  \left(  H_{a}+H_{b}%
+H_{c}\right)  +\dot{H}_{a}+\dot{H}_{b}\right)  +\dot{Q}f^{\prime\prime
}\left(  H_{a}+H_{b}-\frac{3}{2}\gamma\right)  +\frac{f}{2}=0, \label{con2.05}%
\end{align}
with the equation of motion for the connection
\begin{equation}
3\gamma\dot{Q}^{2}f^{\prime\prime\prime}+f^{\prime\prime}\left(  \left(
H_{a}+H_{b}+H_{c}\right)  \dot{Q}+\ddot{Q}\right)  =0. \label{con2.06}%
\end{equation}

\subsubsection{Minisuperspace description}

To write the point-like Lagrangian for the field equations of connection
$\Gamma_{2}$ we follow \cite{min} and we define $\phi=f^{\prime}\left(
Q\right) ,~\gamma=\dot{\psi}$ and $V\left(  \phi\right)  =\frac{1}{2}\left(
f-Qf^{\prime}\right) $. The corresponding Lagrangian is
\begin{equation}
L_{\Gamma_{2}}\left(  a,\dot{a},b,\dot{b},c,\dot{\beta_{2}},\phi,\dot{\phi
},\psi,\dot{\psi}\right)  =\dot{\phi}\left(  c\dot{a}\dot{b}+b\dot{a}\dot
{c}+a\dot{b}\dot{c}\right)  +3abc\dot{\psi}\dot{\phi}-abcV\left(  \phi\right)
. \label{con2.07}%
\end{equation}

\subsection{Field equations for connection $\Gamma_{3}$}

For the third connection, namely $\Gamma_{3}$ we calculate the nonmetricity
scalar
\begin{align}
Q\left(  \Gamma_{3}\right)   &  =-2\left(  H_{a}H_{b}+H_{a}H_{c}+H_{b}%
H_{c}\right)  -\left(  \frac{1}{a^{2}}+\frac{1}{b^{2}}+\frac{1}{c^{2}}\right)
\dot{\gamma}\nonumber\\
&  +H_{a}\gamma\left(  \frac{1}{a^{2}}-\frac{1}{b^{2}}-\frac{1}{c^{2}}\right)
+H_{b}\gamma\left(  \frac{1}{b^{2}}-\frac{1}{a^{2}}-\frac{1}{c^{2}}\right)
+H_{c}\gamma\left(  \frac{1}{c^{2}}-\frac{1}{b^{2}}-\frac{1}{a^{2}}\right)  .
\label{con3.01}%
\end{align}

The field equations are derived
\begin{equation}
f^{\prime}\left(  H_{a}H_{b}+H_{a}H_{c}+H_{b}H_{c}\right)  +\frac{1}{2}\left(
f-Qf^{\prime}\right)  +\dot{Q}f^{\prime\prime}\left(  \frac{1}{a^{2}}+\frac
{1}{b^{2}}+\frac{1}{c^{2}}\right)  \gamma=0, \label{con3.02}%
\end{equation}%
\begin{equation}
f^{\prime}\left(  \left(  H_{b}+H_{c}\right)  \left(  H_{a}+H_{b}%
+H_{c}\right)  +\dot{H}_{b}+\dot{H}_{c}\right)  +\dot{Q}f^{\prime\prime
}\left(  H_{b}+H_{c}+\frac{\gamma}{2}\left(  \frac{1}{a^{2}}-\frac{1}{b^{2}%
}-\frac{1}{c^{2}}\right)  \right)  +\frac{f}{2}=0, \label{con3.03}%
\end{equation}%
\begin{equation}
f^{\prime}\left(  \left(  H_{a}+H_{c}\right)  \left(  H_{a}+H_{b}%
+H_{c}\right)  +\dot{H}_{a}+\dot{H}_{c}\right)  +\dot{Q}f^{\prime\prime
}\left(  H_{a}+H_{c}+\frac{\gamma}{2}\left(  \frac{1}{b^{2}}-\frac{1}{a^{2}%
}-\frac{1}{c^{2}}\right)  \right)  +\frac{f}{2}=0, \label{con3.04}%
\end{equation}%
\begin{equation}
f^{\prime}\left(  \left(  H_{a}+H_{b}\right)  \left(  H_{a}+H_{b}%
+H_{c}\right)  +\dot{H}_{a}+\dot{H}_{b}\right)  +\dot{Q}f^{\prime\prime
}\left(  H_{a}+H_{b}+\frac{\gamma}{2}\left(  \frac{1}{c^{2}}-\frac{1}{b^{2}%
}-\frac{1}{a^{2}}\right)  \right)  +\frac{f}{2}=0. \label{con3.05}%
\end{equation}

Finally, the equation of motion for the connection reads%
\begin{equation}
\dot{Q}^{2}f^{\prime\prime\prime}+\ddot{Q}f^{\prime\prime}+\dot{Q}%
f^{\prime\prime}\left(  2\frac{\dot{\gamma}}{\gamma}+\frac{a^{2}b^{2}\left(
H_{a}+H_{b}-H_{c}\right)  +a^{2}c^{2}\left(  H_{a}+H_{c}-H_{b}\right)
+b^{2}c^{2}\left(  H_{b}+H_{c}-H_{a}\right)  }{a^{2}b^{2}+a^{2}c^{2}%
+b^{2}c^{2}}\right)  =0. \label{con3.06}%
\end{equation}

\subsubsection{Minisuperspace description}

Finally, we introduce $\phi=f^{\prime}\left(  Q\right) ,~\gamma= {\dot{\psi}}$
and $V\left(  \phi\right)  =\frac{1}{2}\left(  f-Qf^{\prime}\right)  $
~\cite{min} and the minisuperspace Lagrangian for the field equations of
connection $\Gamma_{3}$ reads%
\begin{equation}
L_{\Gamma_{3}}\left(  a,\dot{a},b,\dot{b},c,\dot{c},\phi,\dot{\phi},\psi
,\dot{\psi}\right)  =\dot{\phi}\left(  c\dot{a}\dot{b}+b\dot{a}\dot{c}%
+a\dot{b}\dot{c}\right)  -\left(  \frac{1}{a^{2}}+\frac{1}{b^{2}}+\frac
{1}{c^{2}}\right)  {\dot{\phi}}{\dot{\psi}}-abcV\left(  \phi\right)  .
\label{con3.07}%
\end{equation}

\section{Kasner (-like) spacetimes}

\label{sec4}

The Kasner universe is an exact solution of GR \cite{kasner1} in the vacuum
for the Bianchi I line element (\ref{b1}). The Kasner metric has three
parameters, namely the Kasner indices, which must satisfy the two so-called
Kasner algebraic relations, so it is a one-parameter family of solutions.
Specifically, the values of the parameters are defined on the real number line
by the intersection of a three-dimensional sphere of radius unity and a plane
in which the sum of those parameters is one. The importance of the Kasner
universe is that it describes the evolution of the Mixmaster universe (Bianchi
IX) close to the singularity. In \cite{coleyb}, it has been shown that the
Kasner metric describes the general Bianchi models asymptotically at
intermediate stages of their evolution and early and late-times dynamics.

For the line element (\ref{b1}) the Kasner solution is described by the
power-law scale factors
\begin{equation}
a(t) =t^{p_{1}},~b(t) =t^{p_{2}}\text{ and ~}c(t) =t^{p_{3}}, \label{ks.01}%
\end{equation}
where the indices $p_{i}=\left(  p_{i},p_{2},p_{3}\right)  $ satisfy the
Kasner's relations%
\begin{align}
p_{1}+p_{2}+p_{3}  &  =1, \quad 
p_{1}^{2}+p_{2}^{2}+p_{3}^{2}   =1. \label{ks.03}%
\end{align}

Kasner-like power-law solutions are spacetimes with power-law scale factors
with generalized Kasner's relations. These solutions have been studied before
in the literature for the analysis of anisotropies in higher spacetime
dimensions and in modified theories of gravity
\cite{DemaretKL,kas1,kas2,kas3,kas4,kas5,kas6,kas7}.

In the following, we investigate the existence of Kasner (-like) solutions in
nonmetricity $f\left(Q\right)$-gravity for the three different connections.

\subsection{Connection $\Gamma_{1}$}

The field equations (\ref{con.00})-(\ref{con.04}) for connection $\Gamma_{1}$
defined in the coincidence gauge is identical to that of teleparallel $f(t)
$-gravity \cite{ks2}. Hence, the results of $f\left(  T\right)  $-gravity hold
and for the $f\left(Q\right)$-gravity. Thus, for the power-law $f\left(  Q\right)  =\left(  -Q\right)  ^{n}$ model,
the power-law scale factors (\ref{ks.01}) solve the field equations when the
indices $p_{1},~p_{2}$ and $p_{3}$ satisfy the modified Kasner relations%
\begin{align}
p_{1}+p_{2}+p_{3}  &  =2n-1,\\
p_{1}^{2}+p_{2}^{2}+p_{3}^{2}  &  =\left(  2n-1\right)  ^{2}, 
\end{align}
for $n>0$, or%
\begin{align}
p_{1}+p_{2}+p_{3}  &  =\alpha, \quad 
p_{1}^{2}+p_{2}^{2}+p_{3}^{2}  =\alpha^{2}%
\end{align}
for $n>1$.

Indeed, the Kasner solution is recovered when $n=1$, or for $n>1$ and
$\alpha=1$. Furthermore, we remark that one of the resonances always has a
different sign from the others, which means that the chaotic dynamic behavior
near the singularity in the Mixmaster universe is similar to that of GR.

We proceed with the investigation of the existence of Kasner (-like) solutions
for the other two connections, where for simplicity of the calculations, we
work, without loss of generality, in the scalar field description approach.

\subsection{Connection $\Gamma_{2}$}

We assume the scale factors' power-law solution (\ref{ks.01}). By
replacing in the gravitational field equations, we find
\begin{align}
&  \frac{p_{1}p_{2}+p_{1}p_{3}+p_{2}p_{3}}{3t^{2}}\phi+\frac{3}{2}\gamma
\dot{\phi}+V\left(  \phi\right)  =0,\\
&  t\left(  3t\gamma-2\left(  p_{2}+p_{3}\right)  \dot{\phi}-2\left(  \left(
p_{2}-1\right)  p_{2}+p_{3}\left(  p_{3}-1\right)  +p_{3}^{2}\right)  \right)
\phi-2t^{2}V\left(  \phi\right)  =0,\\
&  t\left(  3t\gamma-2\left(  p_{1}+p_{3}\right)  \dot{\phi}-2\left(  \left(
p_{1}-1\right)  p_{1}+p_{3}\left(  p_{1}-1\right)  +p_{3}^{2}\right)  \right)
\phi-2t^{2}V\left(  \phi\right)  =0,\\
&  t\left(  3t\gamma-2\left(  p_{1}+p_{2}\right)  \dot{\phi}-2\left(  \left(
p_{1}-1\right)  p_{1}+p_{2}\left(  p_{1}-1\right)  +p_{2}^{2}\right)  \right)
\phi-2t^{2}V\left(  \phi\right)  =0,
\end{align}
and%
\begin{equation}
t\ddot{\phi}+\left(  p_{1}+p_{2}+p_{3}\right)  \dot{\phi}=0.
\end{equation}

Thus, the solution of the latter system is
\begin{equation}
V\left(  \phi\right)  =0~,~\gamma(t) =-\frac{2\left(  p_{1}p_{2}+p_{1}%
p_{3}+p_{2}p_{3}\right)  }{3\left(  p_{1}+p_{2}+p_{3}\right)  }\frac{1}%
{t}~,~\phi(t) =\phi_{1}t^{1-p_{1}-p_{2}-p_{3}}.
\end{equation}

By replacing in (\ref{con2.01}) we find
\begin{equation}
V_{,\phi}=Q=0.
\end{equation}

Thus, from the definition of the scalar field $\phi=f^{\prime}\left(
Q\right)  $, and of the potential $V\left(  \phi\right)  =f^{\prime}\left(
Q\right)  Q-f\left(  Q\right)  $ we end with the algebraic equation
\begin{equation}
\phi_{1}=0. \label{cn2}%
\end{equation}

We conclude the power-law solution (\ref{ks.01}), which solves the field
equations for the second connection for any indices $p_{1}$, $p_{2}$ and
$p_{3}$ when $f\left(  Q\rightarrow0\right)  =0$ and $f^{\prime}\left(
Q\rightarrow0\right)  =0$. That is, $f\left(Q\right)$ can be any function
in which $Q=0$ is a zero and a stationary point, that is, $f^{\prime}\left(
Q\rightarrow0\right)  =0$. We remark that there is not any Kasner (-like)
relation because of the function $\gamma(t) $.

On the other hand, in the case of the Kasner universe, that is, the Kasner
relations (\ref{ks.03}) are satisfied, we find that $\gamma(t)
=\frac{\gamma_{0}}{t}$, and $Q=0$, $f\left(  Q\rightarrow0\right)  =0$.

\subsection{Connection $\Gamma_{3}$}

From the field equations of the third connection and for the scale factors
(\ref{ks.01}) we find%
\begin{align}
0  &  =t\left(  t^{2\left(  p_{1}+p_{2}\right)  }+t^{2\left(  p_{1}%
+p_{3}\right)  }+t^{2\left(  p_{2}+p_{3}\right)  }\right)  \left(  \ddot{\phi
}\gamma+2\dot{\gamma}\dot{\phi}\right) \nonumber\\
&  +\left(  p_{1}+p_{2}-p_{3}\right)  t^{2\left(  p_{1}+p_{2}\right)
}+\left(  p_{1}-p_{2}+p_{3}\right)  t^{2\left(  p_{1}+p_{3}\right)  }+\left(
-p_{1}+p_{2}+p_{3}\right)  t^{2\left(  p_{2}+p_{3}\right)  },\\
0  &  =2\left(  p_{1}p_{2}+p_{1}p_{3}+p_{2}p_{3}\right)  t^{2\left(
p_{1}+p_{2}+p_{3}\right)  }\phi\\
&  +t^{2}\left(  t^{2\left(  p_{1}+p_{2}\right)  }+t^{2\left(  p_{1}%
+p_{3}\right)  }+t^{2\left(  p_{2}+p_{3}\right)  }\right)  \gamma\dot{\phi
}+t^{2\left(  1+p_{1}+p_{2}+p_{3}\right)  }V\left(  \phi\right),\\
0  &  =\left(  p_{3}^{2}+p_{3}\left(  p_{2}-1\right)  +p_{2}\left(
p_{2}-1\right)  \right)  t^{2\left(  p_{1}+p_{2}+p_{3}\right)  }%
\phi-t^{2\left(  1+p_{1}+p_{2}+p_{3}\right)  }V\left(  \phi\right) \nonumber\\
&  +t\left(  2\left(  p_{1}+p_{2}\right)  t^{2\left(  p_{1}+p_{2}%
+p_{3}\right)  }+t\left(  t^{2\left(  p_{1}+p_{2}\right)  }+t^{2\left(
p_{1}+p_{3}\right)  }-t^{2\left(  p_{2}+p_{3}\right)  }\right)  \gamma\right)
\dot{\phi},\\
0  &  =\left(  p_{3}^{2}+p_{3}\left(  p_{1}-1\right)  +p_{1}\left(
p_{1}-1\right)  \right)  t^{2\left(  p_{1}+p_{2}+p_{3}\right)  }%
\phi-t^{2\left(  1+p_{1}+p_{2}+p_{3}\right)  }V\left(  \phi\right) \nonumber\\
&  +t\left(  2\left(  p_{1}+p_{2}\right)  t^{2\left(  p_{1}+p_{2}%
+p_{3}\right)  }+t\left(  t^{2\left(  p_{1}+p_{2}\right)  }-t^{2\left(
p_{1}+p_{3}\right)  }+t^{2\left(  p_{2}+p_{3}\right)  }\right)  \gamma\right)
\dot{\phi},\\
0  &  =\left(  p_{1}^{2}+p_{1}\left(  p_{2}-1\right)  +p_{2}\left(
p_{2}-1\right)  \right)  t^{2\left(  p_{1}+p_{2}+p_{3}\right)  }%
\phi-t^{2\left(  1+p_{1}+p_{2}+p_{3}\right)  }V\left(  \phi\right) \nonumber\\
&  +t\left(  2\left(  p_{1}+p_{2}\right)  t^{2\left(  p_{1}+p_{2}%
+p_{3}\right)  }+t\left(  t^{2\left(  p_{1}+p_{2}\right)  }+t^{2\left(
p_{1}+p_{3}\right)  }+t^{2\left(  p_{2}+p_{3}\right)  }\right)  \gamma\right)
\dot{\phi}.
\end{align}

Therefore the only solution where $\gamma\neq0$, it is that with $V\left(
\phi\right)  =0,$ $\phi=0$, that is, $Q=0$, $f\left(  Q\rightarrow0\right)
=0$ and $f^{\prime}\left(  Q\rightarrow0\right)  =0$ in which from the
definition of the nonmetricity scalar (\ref{con3.01}) we calculate%
\begin{align}
\gamma(t)  &  =\gamma_{0}\frac{t^{p_{1}+p_{2}+p_{3}}}{t^{2\left(  p_{1}%
+p_{2}\right)  }+t^{2\left(  p_{1}+p_{3}\right)  }+t^{2\left(  p_{2}%
+p_{3}\right)  }}\nonumber\\
&  +\frac{2\left(  p_{1}p_{2}+p_{1}p_{3}+p_{2}p_{3}\right)  }{1-p_{1}%
-p_{2}-p_{3}}\frac{t^{1+2\left(  p_{1}+p_{2}+p_{3}\right)  }}{\left(
t^{2\left(  p_{1}+p_{2}\right)  }+t^{2\left(  p_{1}+p_{3}\right)
}+t^{2\left(  p_{2}+p_{3}\right)  }\right)  }. %
\end{align}

Last but not least, in the case of the Kasner solution, the connection is
calculated%
\begin{equation}
\gamma(t) =\gamma_{0}\frac{t}{t^{2\left(  p_{1}+p_{2}\right)  }+t^{2\left(
p_{1}+p_{3}\right)  }+t^{2\left(  p_{2}+p_{3}\right)  }}.
\end{equation}

\section{Phase-space analysis}

\label{sec5}

In this Section, we use the dimensionless variables and investigate
the stationary points for the field equations. Such analysis provides 
essential information regarding the cosmological history and the model's fate.

\subsection{Connection $\Gamma_{1}$}%
We consider the point-like action
\begin{equation}
\mathcal{L}_{1}\left( a,\dot{a},\beta_{1},\dot{\beta_{1}},c,\dot{\beta_{2}%
},\phi\right)  =\frac{3 a \phi\dot{a}^{2}}{N}-a^{3} \left( \frac{3 \phi\left(
\dot{\beta_{1}}^{2}+ \dot{\beta_{2}}^{2}\right) }{4 N}-N V(\phi)\right) .
\end{equation}

Taking variations of $\mathcal{L}_{1}$ with respect to $\{N, a, \beta_{1},
\beta_{2}, \phi\}$, and setting $N=1$ after variations we obtain the system
\begin{align}
&  a^{2} \left( 3 \phi\left( \dot{\beta_{1}}^{2}+\dot{\beta_{2}}^{2}\right) +4
V(\phi)\right) -12 \phi\dot{a}^{2} =0,\\
&  4 \phi\dot{a}^{2}+a \left( 8 \phi\ddot{a}+8 \dot{a} \dot{\phi}+a \left( 3
\phi\left( \dot{\beta_{1}}^{2}+\dot{\beta_{2}}^{2}\right) -4 V(\phi)\right)
\right) =0,\\
&  \dot{\beta_{1}} \left( 3 \phi\dot{a}+a \dot{\phi}\right) +a \phi\ddot
{\beta_{1}}=0,\\
&  \dot{\beta_{2}} \left( 3 \phi\dot{a}+a \dot{\phi}\right) +a \phi\ddot
{\beta_{2}}=0,\\
&  a^{2} \left( 3 \left( \dot{\beta_{1}}^{2}+\dot{\beta_{2}}^{2}\right) -4
V^{\prime}(\phi)\right) -12 \dot{a}^{2} = 0 .
\end{align}
Now we define
\begin{equation}
H = \frac{\dot{a}}{a}, \quad\sigma_{1} = \frac{\dot{\beta_{1}}}{2},
\quad\sigma_{2} = \frac{\dot{\beta_{2}}}{2}.
\end{equation}
Thus, we investigate the system
\begin{align}
&  \dot{H} = \frac{-3 \phi\left( H^{2}+\sigma_{1}^{2}+\sigma_{2}^{2}\right) -2
H \dot{\phi}+V(\phi)}{2 \phi},\\
&  \dot{\sigma}_{1} = -{\sigma_{1} \left( 3 H \phi+\dot{\phi}\right)}/{\phi},\\
&  \dot{\sigma}_{2} = -{\sigma_{2} \left( 3 H \phi+\dot{\phi}\right)}/{\phi},
\\
&  V^{\prime}(\phi)= 3 \left( -H^{2}+\sigma_{1}^{2}+\sigma_{2}^{2}\right) .
\end{align}
with restriction
\begin{equation}
3 \phi\left( -H^{2}+\sigma_{1}^{2}+\sigma_{2}^{2}\right) +V(\phi)=0.
\end{equation}

Now, we define the variables
\begin{equation}
y = \frac{V(\phi)}{3 H^{2} \phi}, \quad\quad\Sigma_{1} = \frac{\sigma_{1}}{H},
\quad\Sigma_{2} = \frac{\sigma_{2}}{H},
\end{equation}
which satisfy
\begin{equation}
\Sigma_{1}^{2} +\Sigma_{2}^{2}+y=1.
\end{equation}
This expression can be used as a definition of $y$.

We have the equations
\begin{align}
&  \Sigma_{1}^{\prime}=\frac{3}{2}\Sigma_{1}\left(  \Sigma_{1}^{2}+\Sigma
_{2}^{2}-y-1\right), \\
&  \Sigma_{2}^{\prime}=\frac{3}{2}\Sigma_{2}\left(  \Sigma_{1}^{2}+\Sigma
_{2}^{2}-y-1\right), \\
&  y^{\prime}=-6\left(  \Sigma_{1}^{2}+\Sigma_{2}^{2}-1\right)  \left(
\Sigma_{1}^{2}+\Sigma_{2}^{2}\right),
\end{align}
where we have defined the new derivative
\begin{equation}
y^{\prime}\equiv\frac{1}{H}\dot{y}.
\end{equation}

Finally, we investigate the reduced system
\begin{align}
&  \Sigma_{1}^{\prime} = 3 \Sigma_{1}\left( \Sigma_{1}^{2}+\Sigma_{2}%
^{2}-1\right) ,\label{0eqS1}\\
&  \Sigma_{2}^{\prime} = 3 \Sigma_{2} \left( \Sigma_{1}^{2}+\Sigma_{2}%
^{2}-1\right) .\label{0eqS2}%
\end{align}

The equilibrium points/lines of points/ surfaces of
the system \eqref{0eqS1} and \eqref{0eqS2} are the following.

\begin{enumerate}
\item The circle 
$$A: \Sigma_{2}^2 +\Sigma_{1}^{2}=1$$ and $y=0$ with eigenvalues $6, 0$. It is unstable. 
\item The point $C: (0,0)$ and $y=1$ with eigenvalues $-3,-3$ that is a sink.
\end{enumerate}

\begin{table}
\centering
\begin{tabular}
[c]{|c|c|c|c|c|c|}\hline
Label & Location & $y$ & $k_{1}$ & $k_{2}$ &
Behavior\\\hline
$A$ & $\Sigma_{2}^2 +\Sigma_{1}^{2}=1$ & $0$ & $6$ & $0$ & unstable\\
$B$ & $\Sigma_{1}=0, \Sigma_{2}=0$ & $1$ & $-3$ & $-3$ & sink\\\hline
\end{tabular}
\caption{Equilibrium points/lines of points/ surfaces of the system
\eqref{0eqS1} and \eqref{0eqS2}.}%
\label{tab:TAB0}%
\end{table}

The equilibrium points/lines of points/ surfaces of the system \eqref{0eqS1}
and \eqref{0eqS2} are listed on Table \ref{tab:TAB0}. The lines of points $A$ represent Kasner solutions. The point $B$ is the attractor and corresponds
to a potential-dominated solution.

\subsection{Connection $\Gamma_{2}$}

We consider the point-like action
\begin{equation}
\mathcal{L}_{2}(N,a,\dot{a},b,\dot{b},c,\dot{c},\phi,\dot{\phi},\psi,\dot
{\psi}):=\frac{3a\phi\dot{a}^{2}}{N}-a^{3}\left(  \frac{3\phi\left(
\dot{\beta_{1}}^{2}+\dot{\beta_{2}}^{2}\right)  }{4N}-NV(\phi)+\frac
{3\dot{\psi}\dot{\phi}}{2N}\right),
\end{equation}
where $\gamma=\dot{\psi}$. Taking variations of $\mathcal{L}_{2}$ with respect
to $\{N,a,\beta_{1},\beta_{2},\phi,\psi\}$, and setting $N=1$ after variations
we obtain the system
\begin{align}
&  a^{3}\left(  3\phi\left(  \dot{\beta_{1}}^{2}+\dot{\beta_{2}}^{2}\right)
+4V(\phi)+6\dot{\psi}\dot{\phi}\right)  -12a\phi\dot{a}^{2}=0,\\
&  -4\phi\dot{a}^{2}-8a\left(  \phi\ddot{a}+\dot{a}\dot{\phi}\right)
+a^{2}\left(  -3\phi\left(  \dot{\beta_{1}}^{2}+\dot{\beta_{2}}^{2}\right)
+4V(\phi)-6\dot{\psi}\dot{\phi}\right)  =0,\\
&  \phi\left(  3\dot{a}\dot{\beta_{1}}+a\ddot{\beta_{1}}\right)  +a\dot
{\beta_{1}}\dot{\phi}=0,\\
&  \phi\left(  3\dot{a}\dot{\beta_{2}}+a\ddot{\beta_{2}}\right)  +a\dot
{\beta_{2}}\dot{\phi}=0,\\
&  18a\dot{a}\dot{\psi}+12\dot{a}^{2}+a^{2}\left(  4V^{\prime}(\phi)-3\left(
\dot{\beta_{1}}^{2}+\dot{\beta_{2}}^{2}-2\ddot{\psi}\right)  \right)  =0,\\
&  3\dot{a}\dot{\phi}+a\ddot{\phi}=0
\end{align}
Now we define
\begin{equation}
H=\frac{\dot{a}}{a},\quad\sigma_{1}=\frac{\dot{\beta_{1}}}{2},\quad\sigma
_{2}=\frac{\dot{\beta_{2}}}{2}.
\end{equation}
Thus, we investigate the system
\begin{align}
&  \dot{H}=-\frac{4H\dot{\phi}+6H^{2}\phi+6\phi\left(  \sigma_{1}^{2}%
+\sigma_{2}^{2}\right)  -2V(\phi)+3\dot{\psi}\dot{\phi}}{4\phi},\\
&  \dot{\sigma_{1}}=-{\sigma_{1}\left(  3H\phi+\dot{\phi}\right)}/{\phi},\\
&  \dot{\sigma_{2}}=-{\sigma_{2}\left(  3H\phi+\dot{\phi}\right)}/{\phi},\\
&  \ddot{\phi}=-3H\dot{\phi},\\
&  \ddot{\psi}=-3H\dot{\psi}-2H^{2}+2\left(  \sigma_{1}^{2}+\sigma_{2}%
^{2}\right)  -\frac{2}{3}V^{\prime}(\phi),
\end{align}
with restriction
\begin{equation}
6\phi\left(  -H^{2}+\sigma_{1}^{2}+\sigma_{2}^{2}\right)  +2V(\phi)+3\dot
{\psi}\dot{\phi}=0.
\end{equation}
Defining the new variables
\[
x=\frac{\dot{\phi}}{H\phi},\quad y=\frac{V(\phi)}{3H^{2}\phi},\quad
z=\frac{\dot{\psi}}{2H},\quad\Sigma_{1}=\frac{\sigma_{1}}{H},\quad\Sigma
_{2}=\frac{\sigma_{2}}{H},
\]
with inverse
\[
V(\phi)=3H^{2}y\phi,\quad\sigma_{1}=H\Sigma_{1},\quad\sigma_{2}=H\Sigma
_{2},\quad\dot{\phi}Hx\phi,\quad\dot{\psi}=2Hz
\]
and the auxiliary variables
\begin{align}
&  K=\frac{\phi^{2}V^{\prime\prime}(\phi)}{V(\phi)}-\frac{\phi^{2}V^{\prime2}%
}{V(\phi)^{2}}+\frac{\phi V^{\prime}(\phi)}{V(\phi)},  \quad \lambda=\frac{\phi V^{\prime}(\phi)}{V(\phi)},
\end{align}
with inverse
\begin{align}
&  V^{\prime}(\phi)=\frac{\lambda V(\phi)}{\phi}, \quad  V^{\prime\prime}(\phi)=\frac{\left(  K+\lambda^{2}-\lambda\right)  V(\phi
)}{\phi^{2}},
\end{align}
we obtain the dynamical system
\begin{align}
&  \Sigma_{1}^{\prime}=\frac{3}{2}\Sigma_{1}\left(  \Sigma_{1}^{2}+\Sigma
_{2}^{2}+xz-y-1\right), \\
&  \Sigma_{2}^{\prime}=\frac{3}{2}\Sigma_{2}\left(  \Sigma_{1}^{2}+\Sigma
_{2}^{2}+xz-y-1\right), \\
&  x^{\prime}=\frac{3}{2}x\left(  \Sigma_{1}^{2}+\Sigma_{2}^{2}+xz-y-1\right)
,\\
&  y^{\prime}=y\left(  x(\lambda+3z+1)+3\left(  \Sigma_{1}^{2}+\Sigma_{2}%
^{2}-y+1\right)  \right), \\
&  z^{\prime}=\frac{1}{2}\left(  (3z+2)\left(  \Sigma_{1}^{2}+\Sigma_{2}%
^{2}+xz-1\right)  -y(2\lambda+3z)\right), \\
&  \lambda^{\prime}=Kx, \label{evol-lambda}%
\end{align}
where we have defined the new derivative
\begin{equation}
x^{\prime}\equiv\frac{1}{H}\dot{x}
\end{equation}
and it can be assume that $K$ is an explicit function of $\lambda$,
$K=K(\lambda)$. Therefore, we obtain a closed dynamical system.

It satisfied the restriction
\begin{equation}
\Sigma_{1}^{2}+\Sigma_{2}^{2}+xz+y=1.
\end{equation}
That is a conservation law for the flow. Therefore, we study the reduced
system
\begin{align}
&  \Sigma_{1}^{\prime}=3\Sigma_{1}\left(  \Sigma_{1}^{2}+\Sigma_{2}%
^{2}+xz-1\right), \label{syst1}\\
&  \Sigma_{2}^{\prime}=3\Sigma_{2}\left(  \Sigma_{1}^{2}+\Sigma_{2}%
^{2}+xz-1\right), \label{syst2}\\
&  x^{\prime}=3x\left(  \Sigma_{1}^{2}+\Sigma_{2}^{2}+xz-1\right)
,\label{syst3}\\
&  z^{\prime}=(\lambda+3z+1)\left(  \Sigma_{1}^{2}+\Sigma_{2}^{2}+xz-1\right)
,\label{syst4}\\
&  \lambda^{\prime}=Kx.\label{syst5}%
\end{align}
In general, $\lambda$ is a function of the scalar field with its evolution
equation \eqref{evol-lambda}. For simplicity, we investigate the simple case
$\lambda$ is a constant. That is, $K\equiv0$ and
\begin{equation}
V(\phi)=c_{1}\phi^{\lambda},\quad f(Q)=Qf^{\prime}(Q)+2c_{1}f^{\prime}(Q)^{\lambda}.
\end{equation}
The second is the Clairaut equation, which has a non-trivial ($f\left(
Q\right)  \neq a_{1}+a_{2}Q$) powerlaw solution. It follows
\begin{equation}
f(Q)=\alpha(-Q)^{p},\quad p=\frac{\lambda}{\lambda-1},\quad\alpha=2^{\frac
{1}{1-\lambda}}(\lambda-1)(-\lambda)^{\frac{\lambda}{1-\lambda}}(-c_{1}%
){}^{\frac{1}{1-\lambda}}.
\end{equation}

\begin{table}
\centering
\begin{tabular}
[c]{|c|c|c|c|c|c|c|}\hline
Label & Location & $k_{1}$ & $k_{2}$  & $k_{3}$ & $k_{4}$ &
Behavior\\\hline
$A$ & $\Sigma_{1}^{2}+\Sigma_{2}^{2}=1,x=0$ & $6$ & $0$ & $0$ & $0$ & unstable\\
$B$ & $\Sigma_{1}=0,\Sigma_{2}=0,x=0,z=-\frac{1}{3}(1+\lambda)$ & $-3$ & $-3$ & $-3$ & $-3$ & sink\\
$C$ & $\Sigma_{1}^{2}+\Sigma_{2}^{2}+xz=1,\quad x\neq0\quad z\neq0$ & $0$ & $0$ & $0$ & $\lambda x+x+6$ & unstable\\\hline
\end{tabular}
\caption{Equilibrium points/lines of points/ surfaces of the system
\eqref{syst1}, \eqref{syst2}, \eqref{syst3} and 
\eqref{syst4} for $\lambda$ constant.}%
\label{tab:TAB1}%
\end{table}

In Table \ref{tab:TAB1} are presented the equilibrium points/lines of points/ surfaces of the system
\eqref{syst1}, \eqref{syst2}, \eqref{syst3} and 
\eqref{syst4} for $\lambda$ constant.

The equilibrium sets are:

\begin{enumerate}
\item The circle
\[
A:\Sigma_{1}^{2}+\Sigma_{2}^{2}=1,x=0.
\]
It corresponds to the family Kasner's solutions derived before. The
eigenvalues are ${6,0,0,0}$.

\item The equilibrium points
\[
B:\Sigma_{1}=0,\Sigma_{2}=0,x=0,z=-\frac{1}{3}(1+\lambda).
\]
The eigenvalues are ${-3,-3,-3,-3}$. It is a sink.\ The asymptotic solution is
that of the de Sitter universe.

\item The surface
\[
C:\Sigma_{1}^{2}+\Sigma_{2}^{2}+xz=1,\quad x\neq0\quad z\neq0.
\]
It corresponds to $y=0$. The eigenvalues are $\{0,0,0,\lambda x+x+6\}$. The
asymptotic solution describes Kasner-like universes.

\end{enumerate}

\subsection{Connection $\Gamma_{3}$}

In this example, we have the point-like Lagrangian
\begin{align}
&  \mathcal{L}_{3}(N,a,\dot{a},\beta_{1},\dot{\beta_{1}},\beta_{2},\dot
{\beta_{2}},\phi,\dot{\phi},\Psi,\dot{\Psi}):=\nonumber\\
&  \frac{3a\phi\dot{a}^{2}}{N}-\frac{3a^{3}\phi\dot{\beta_{1}}^{2}}{4N}%
-\frac{aNe^{\beta_{1}-\sqrt{3}\beta_{2}}\dot{\phi}}{2\dot{\Psi}}%
-\frac{aNe^{\beta_{1}+\sqrt{3}\beta_{2}}\dot{\phi}}{2\dot{\Psi}}%
-\frac{ae^{-2\beta_{1}}N\dot{\phi}}{2\dot{\Psi}}-\frac{3a^{3}\phi\dot
{\beta_{2}}^{2}}{4N}-a^{3}NV(\phi)
\end{align}
where $\gamma=1/\dot{\Psi}$. 
Taking variations of $\mathcal{L}_{2}$ with
respect to $\{N,a,\beta_{1},\beta_{2},\phi,\Psi\}$, and setting $N=1$ after
variations, we obtain the system
\begin{align}
&  \left(  3\phi\left(  \dot{\beta_{1}}^{2}+\dot{\beta_{2}}^{2}\right)
-4V(\phi)\right)  a^{2}-12\phi\dot{a}^{2}-\frac{2e^{-2\beta_{1}}\left(
1+e^{3\beta_{1}-\sqrt{3}\beta_{2}}+e^{3\beta_{1}+\sqrt{3}\beta_{2}}\right)
\dot{\phi}}{\dot{\Psi}}=0,\\
&  -3\left(  4V(\phi)+3\phi\left(  \dot{\beta_{1}}^{2}+\dot{\beta_{2}}%
^{2}\right)  \right)  a^{2}-24\left(  \dot{a}\dot{\phi}+\phi\ddot{a}\right)
a\nonumber\\
&  -2\left(  6\phi\dot{a}^{2}+\frac{e^{-2\beta_{1}}\left(  1+e^{3\beta
_{1}-\sqrt{3}\beta_{2}}+e^{3\beta_{1}+\sqrt{3}\beta_{2}}\right)  \dot{\phi}%
}{\dot{\Psi}}\right)  =0,
\end{align}
\begin{align}
&  3\left(  \dot{\beta_{1}}\dot{\phi}+\phi\ddot{\beta_{1}}\right)  a^{2}%
+9\phi\dot{a}\dot{\beta_{1}}a-\frac{e^{-2\beta_{1}-\sqrt{3}\beta_{2}}\left(
e^{3\beta_{1}}-2e^{\sqrt{3}\beta_{2}}+e^{3\beta_{1}+2\sqrt{3}\beta_{2}%
}\right)  \dot{\phi}}{\dot{\Psi}}=0,\\
&  3\left(  \dot{\beta_{2}}\dot{\phi}+\phi\ddot{\beta_{2}}\right)  a^{2}%
+9\phi\dot{a}\dot{\beta_{2}}a-\frac{\sqrt{3}e^{\beta_{1}-\sqrt{3}\beta_{2}%
}\left(  -1+e^{2\sqrt{3}\beta_{2}}\right)  \dot{\phi}}{\dot{\Psi}}=0,\\
&  \frac{2e^{-2\beta_{1}-\sqrt{3}\beta_{2}}}{\dot{\Psi}^{2}}\Bigg(\left(
e^{3\beta_{1}}+e^{\sqrt{3}\beta_{2}}+e^{3\beta_{1}+2\sqrt{3}\beta_{2}}\right)
\dot{a}\dot{\Psi}\nonumber\\
&  -a\Big(2e^{2\beta_{1}+\sqrt{3}\beta_{2}}a^{2}V^{\prime}(\phi)\dot{\Psi}%
^{2}-\left(  e^{3\beta_{1}}-2e^{\sqrt{3}\beta_{2}}+e^{3\beta_{1}+2\sqrt
{3}\beta_{2}}\right)  \dot{\beta_{1}}\dot{\Psi}\nonumber\\
&  -\sqrt{3}e^{3\beta_{1}}\left(  -1+e^{2\sqrt{3}\beta_{2}}\right)  \dot
{\beta_{2}}\dot{\Psi}+e^{3\beta_{1}}\ddot{\Psi}+e^{\sqrt{3}\beta_{2}}%
\ddot{\Psi}+e^{3\beta_{1}+2\sqrt{3}\beta_{2}}\ddot{\Psi}\Big)\Bigg)\nonumber\\
&  -3a\left(  a^{2}\left(  \dot{\beta_{1}}^{2}+\dot{\beta_{2}}^{2}\right)
-4\dot{a}^{2}\right)  =0,\\
&  -\left(  e^{3\beta_{1}}+e^{\sqrt{3}\beta_{2}}+e^{3\beta_{1}+2\sqrt{3}%
\beta_{2}}\right)  \dot{a}\dot{\phi}\dot{\Psi}\nonumber\\
&  -a\Bigg(\left(  e^{3\beta_{1}}-2e^{\sqrt{3}\beta_{2}}+e^{3\beta_{1}%
+2\sqrt{3}\beta_{2}}\right)  \dot{\beta_{1}}\dot{\phi}\dot{\Psi}+\sqrt
{3}e^{3\beta_{1}}\left(  -1+e^{2\sqrt{3}\beta_{2}}\right)  \dot{\beta_{2}}%
\dot{\phi}\dot{\Psi}\nonumber\\
&  +\left(  e^{3\beta_{1}}+e^{\sqrt{3}\beta_{2}}+e^{3\beta_{1}+2\sqrt{3}%
\beta_{2}}\right)  \left(  \dot{\Psi}\ddot{\phi}-2\dot{\phi}\ddot{\Psi
}\right)  \Bigg)=0.
\end{align}
Now we define
\begin{equation}
H=\frac{\dot{a}}{a},\quad\sigma_{1}=\frac{\dot{\beta_{1}}}{2},\quad\sigma
_{2}=\frac{\dot{\beta_{2}}}{2}.
\end{equation}
Thus, we investigate the system
\begin{align}
&  \dot{H}=\frac{\dot{\phi}\left(  -6a^{2}H\dot{\Psi}+e^{\beta_{1}-\sqrt
{3}\beta_{2}}+e^{\beta_{1}+\sqrt{3}\beta_{2}}+e^{-2\beta_{1}}\right)  }%
{6a^{2}\phi\dot{\Psi}}-3\left(  \sigma_{1}^{2}+\sigma_{2}^{2}\right), \\
&  \dot{\sigma_{1}}=\frac{e^{\beta_{1}-\sqrt{3}\beta_{2}}\dot{\phi}}%
{6a^{2}\phi\dot{\Psi}}+\frac{e^{\beta_{1}+\sqrt{3}\beta_{2}}\dot{\phi}}%
{6a^{2}\phi\dot{\Psi}}-\frac{e^{-2\beta_{1}}\dot{\phi}}{3a^{2}\phi\dot{\Psi}%
}-3H\sigma_{1}-\frac{\sigma_{1}\dot{\phi}}{\phi},\\
&  \dot{\sigma_{2}}=-\frac{e^{\beta_{1}-\sqrt{3}\beta_{2}}\dot{\phi}}%
{2\sqrt{3}a^{2}\phi\dot{\Psi}}+\frac{e^{\beta_{1}+\sqrt{3}\beta_{2}}\dot{\phi
}}{2\sqrt{3}a^{2}\phi\dot{\Psi}}-3H\sigma_{2}-\frac{\sigma_{2}\dot{\phi}}%
{\phi},\\
&  \ddot{\phi}=\frac{\dot{\phi}}{e^{3\beta_{1}+2\sqrt{3}\beta_{2}}%
+e^{3\beta_{1}}+e^{\sqrt{3}\beta_{2}}}\Bigg(e^{\sqrt{3}\beta_{2}}\left(
4a^{2}e^{2\beta_{1}}\dot{\Psi}\left(  3H^{2}-3\left(  \sigma_{1}^{2}%
+\sigma_{2}^{2}\right)  -V^{\prime}(\phi)\right)  +H-4\sigma_{1}\right)
\nonumber\\
&  +e^{3\beta_{1}+2\sqrt{3}\beta_{2}}\left(  H+2\left(  \sigma_{1}+\sqrt
{3}\sigma_{2}\right)  \right)  +e^{3\beta_{1}}\left(  H+2\sigma_{1}-2\sqrt
{3}\sigma_{2}\right)  \Bigg),
\end{align}
\begin{align}
&  \ddot{\Psi}=\frac{\dot{\Psi}}{e^{3\beta_{1}+2\sqrt{3}\beta_{2}}%
+e^{3\beta_{1}}+e^{\sqrt{3}\beta_{2}}}\Bigg(e^{\sqrt{3}\beta_{2}}\left(
2a^{2}e^{2\beta_{1}}\dot{\Psi}\left(  3H^{2}-3\left(  \sigma_{1}^{2}%
+\sigma_{2}^{2}\right)  -V^{\prime}(\phi)\right)  +H-4\sigma_{1}\right)
\nonumber\\
&  +e^{3\beta_{1}+2\sqrt{3}\beta_{2}}\left(  H+2\left(  \sigma_{1}+\sqrt
{3}\sigma_{2}\right)  \right)  +e^{3\beta_{1}}\left(  H+2\sigma_{1}-2\sqrt
{3}\sigma_{2}\right)  \Bigg),
\end{align}
with restriction
\[
V(\phi)=-\frac{e^{\beta_{1}-\sqrt{3}\beta_{2}}\dot{\phi}}{2a^{2}\dot{\Psi}%
}-\frac{e^{\beta_{1}+\sqrt{3}\beta_{2}}\dot{\phi}}{2a^{2}\dot{\Psi}}%
-\frac{e^{-2\beta_{1}}\dot{\phi}}{2a^{2}\dot{\Psi}}-3H^{2}\phi+3\sigma_{1}%
^{2}\phi+3\sigma_{2}^{2}\phi.
\]

We define the new variables
\begin{align}
&  x=\frac{\dot{\phi}}{H\phi},\quad y=\frac{V(\phi)}{3H^{2}\phi},~z=\frac
{1}{6a^{2}H\dot{\psi}},\nonumber\\
&  u=e^{\beta_{1}-\sqrt{3}\beta_{2}},\quad v=e^{\beta_{1}+\sqrt{3}\beta_{2}%
},~\Sigma_{1}=\frac{\sigma_{1}}{H},\quad\Sigma_{2}=\frac{\sigma_{2}}%
{H}.\nonumber
\end{align}
which satisfies the constraint equation
\begin{equation}
1+y-\Sigma_{1}^{2}-\Sigma_{2}^{2}+\frac{x}{z}\left(  u+v+\frac{1}{uv}\right)
=0.
\end{equation}
Defining the time variable
\begin{equation}
x^{\prime}\equiv\frac{1}{H}\dot{x},
\end{equation}
and assuming that $\lambda$ is a constant parameter similar to before and
replace $y$ from the constraint equation, we end with the following dynamical
system
\begin{align}
x^{\prime}  & =-\frac{x^{2}\left(  1+uv\left(  u+v+\lambda z\right)  \right)
}{uvz}\nonumber\\
& +\frac{x\left(  1+\Sigma_{1}\left(  3\Sigma_{1}-4\right)  +3\Sigma^{2}%
+u^{2}v\left(  1+\Sigma_{1}\left(  2+3\Sigma_{1}\right)  -2\sqrt{3}\Sigma
_{2}+3\Sigma_{2}^{2}\right)  \right)  }{1+uv\left(  u+v\right)  }+\nonumber\\
& +\frac{xuv\left(  z\left(  \lambda-2\right)  \left(  1-\Sigma_{1}^{2}%
-\Sigma_{2}^{2}\right)  +v\left(  1+\Sigma_{1}\left(  2+3\Sigma_{1}\right)
+2\sqrt{3}\Sigma_{2}+3\Sigma_{2}^{2}\right)  \right)  }{1+uv\left(
u+v\right)  },
\\
2z^{2}z^{\prime}  & =x\left(  \left(  2+\lambda\right)  z-2\left(
u+v+\frac{1}{uv}\right)  \right)  +\frac{z\left(  8\Sigma_{1}-6\left(
1-\Sigma_{1}^{2}-\Sigma_{2}^{2}\right)  \right)  }{1+uv\left(  u+v\right)
}\nonumber\\
& +\frac{2zu^{2}v\left(  \Sigma_{1}\left(  3\Sigma_{1}-2\right)  -3+2\sqrt
{3}\Sigma_{2}+3\Sigma_{2}^{2}\right)  }{1+uv\left(  u+v\right)  }+\nonumber\\
& +\frac{zuv\left(  \left(  \lambda-2\right)  \left(  1-\Sigma_{1}^{2}%
-\Sigma_{2}^{2}\right)  z-2v\left(  3-2\Sigma_{1}+3\Sigma_{1}^{2}-2\sqrt
{3}\Sigma_{2}+3\Sigma_{2}^{2}\right)  \right)  }{1+uv\left(  u+v\right)  },
\end{align}
\begin{align}
\Sigma_{1}^{\prime} & =-\frac{x\left(  2+\Sigma_{1}+uv\left(  u+v\right)  \left(
\Sigma_{1}-1\right)  \right)  }{uvz}-3\Sigma_{1}\left(  1-\Sigma_{1}%
^{2}-\Sigma_{2}^{2}\right), 
\\
\Sigma_{2}^{^{\prime}}& =-\frac{x\left(  \Sigma_{2}-uv\left(  u+v\right)
\left(  \sqrt{3}-\Sigma_{2}\right)  \right)  }{uvz}-3\Sigma_{2}\left(
1-\Sigma_{1}^{2}-\Sigma_{2}^{2}\right), 
\\
u^{\prime}& =2u\left(  \Sigma_{1}-\sqrt{3}\Sigma_{2}\right), 
\\
v^{\prime} & =2v\left(  \Sigma_{1}+\sqrt{3}\Sigma_{2}\right).
\end{align}

The equilibrium sets are:

\begin{enumerate}
\item The surface
\[
D:x=0,~\Sigma_{1}=0,~\Sigma_{2}=0,~z=\frac{6\left(  1+uv\left(  u+v\right)
\right)  }{\left(  \lambda-2\right)  uv}~,~zuv\neq0.
\]
It corresponds to the isotropic de Sitter universe. The eigenvalues are
${0,0,-5,-3,-3,}\left(  \frac{\left(  \lambda-2\right)  uv}{12\left(
1+uv\left(  u+v\right)  \right)  }\right)  ^{2}$, which means that the point is
a saddle.

\item The point
\[
E:x=0,~\Sigma_{1}=0,~\Sigma_{2}=0,~z=-3~,~u=1~,~v=1.
\]
It corresponds to $y=0$ and describes a homogeneous FLRW universe with a
radiation source. The eigenvalues are $-1+i\sqrt{3}$, $-1+i\sqrt{3}$,
$-1-i\sqrt{3}$, $-1-i\sqrt{3}$, $\frac{18-19\lambda+\sqrt{\lambda\left(
361\lambda-540\right)  -540}}{36}$, $\frac{18-19\lambda+\sqrt{\lambda\left(
361\lambda-540\right)  -540}}{36}$, from where it follows that the point is
attractor for $\frac{18}{19}<\lambda<6$.

\item The equilibrium point
\[
F:x=\frac{20}{2+3\lambda},~\Sigma_{1}=0,~\Sigma_{2}=0,~z=-\frac{12}{\lambda
-2}~,~u=1~,~v=1
\]
describes an isotropic universe with $y=-1-\frac{5\left(  2-\lambda\right)
}{2+3\lambda}$, and the equation of state parameter for the fluid source is
$\frac{14+\lambda}{3\left(  2+3\lambda\right)  }$. \ Finally, the point is an
attractor for $\lambda<-4$.
\end{enumerate}

We conclude that no asymptotic solutions describe anisotropy for the connection $\Gamma_{3}$. Recall that the limit $z\rightarrow0$,
the connection $\Gamma_{3}$ reduced to that of $\Gamma_{1}$; this case is
not new.

\section{Conclusions}

\label{sec6}

Kasner and Kasner-like spacetimes belong to the family of self-similar
solutions within the Bianchi I geometry. Such solutions hold significant
importance in gravitational physics due to their association with similarity
transformations, specifically Homothetic symmetry, which allows mapping
solutions to equivalent ones after appropriately scaling independent variables
\cite{bok1}. Self-similar solutions accurately depict behavior in the
asymptotic limit for more general solutions \cite{as6}.

This analysis explores self-similar solutions for Bianchi I geometries
under symmetric teleparallel $f(Q)$-gravity. Bianchi I geometry can be viewed
as the anisotropic extension of spatially flat FLRW spacetime. Accordingly, we
utilize three families of connections introduced in the case of FLRW spacetime
to define the gravitational theory. Each connection yields distinct sets of
gravitational field equations. We require a self-similar solution and
determine the unknown variables accordingly.

For the first connection, defined in the coincident gauge, we identify Kasner(-like) solutions that extend the Kasner relations. However, the
characteristics of the indices for the exact solution remain unchanged,
indicating a similarity in behavior with the Kasner universe. Conversely, self-similar solutions exist for
the remaining two set of field equations defined by connections in the
noncoincident gauge without Kasner-like constraint equations. That suggests that the chaotic behavior
observed in the Mixmaster universe near the singularity may be absent in
symmetric teleparallel $f\left(Q\right)$-gravity.

Finally, we employed the phase-space analysis to understand the
evolution of anisotropies in this gravitational model. The cosmological evolution was found to depend on the definition of the connection. For all
the connections, the future solution is that of the de Sitter universe.

Recall that in symmetric teleparallel gravity, in the presence of spatial
curvature, the connection that reproduces the field equations is defined in
the noncoincidence gauge \cite{cos2,dm8}. Thus, in future work, we plan to
investigate the evolution of the kinematic variables in the Mixmaster universe and investigate which limit of Bianchi I near the singularity exists.

\begin{acknowledgments}
AP thanks the support of Vicerrectoría de Investigación y Desarrollo Tecnológico (VRIDT) at Universidad Católica del Norte (UCN)  through Resoluci\'{o}n VRIDT No. 096/2022 and
Resoluci\'{o}n VRIDT No. 098/2022. AP thanks ND and the Universidad de La
Frontera for the hospitality provided while part of this work was carried out. VRIDT at UCN 
funded G.L. through Resolución VRIDT No. 026/2023 and Resolución VRIDT No. 027/2023. He acknowledges the support of Núcleo de Investigación Geometría Diferencial y Aplicaciones (Resolución VRIDT No. 096/2022). G. L. also acknowledge the financial support of Proyecto de Investigación Pro Fondecyt Regular 2023 (Resolución VRIDT N°076/2023), Resolución VRIDT N°09/2024 and Agencia Nacional de Investigación y Desarrollo (ANID) through Proyecto Fondecyt Regular 2024,  Folio 1240514, Etapa 2024.

\end{acknowledgments}

\end{document}